\documentclass[12pt,a4paper]{article}

\usepackage[utf8]{inputenc}
\usepackage{graphicx}

\usepackage[a4paper,rmargin=1.7cm,lmargin=1.7cm,tmargin=1.0cm,bmargin=1.0cm,includefoot]{geometry}

\usepackage{array}
\usepackage{epsfig}

\usepackage{cite}

\usepackage[namelimits]{amsmath}
\usepackage{amssymb}
\usepackage{amsfonts}
\usepackage{bbm}

\usepackage{multirow}
\usepackage{color}

\usepackage[T1]{fontenc}      
\usepackage{ae}               
\usepackage{indentfirst}

\usepackage[affil-it]{authblk}

\usepackage{authblk}
\newcommand{\sigmael}{\sigma_\mathrm{el}}
\newcommand{\sigmatot}{\sigma_\mathrm{tot}}

\newcommand{\sigmain}{\sigma_\mathrm{inel}}
\newcommand{\sigmadiff}{\sigma_\mathrm{diff}}

\begin{document}
\title{Updating an empirical analysis on the proton's central opacity and asymptotia}

\author[1]{D. A. Fagundes}
\author[2]{M. J. Menon}
\author[2]{ P. V. R. G. Silva}
\affil[1]{Universidade Federal de Santa Catarina - Campus Blumenau
Rua Pomerode, 710, Salto do Norte, 89065-300
Blumenau, SC, Brazil}
\affil[1]{Instituto de F\'{\i}sica Te\'orica - UNESP
Rua Dr. Bento T. Ferraz, 271, Bloco II, 01140-070
S\~ao Paulo, SP, Brazil}
\affil[2]{Instituto de F\'{\i}sica Gleb Wataghin, Universidade Estadual de Campinas, 
13083-859 Campinas, SP, Brazil}

\date{}

\maketitle

\begin{abstract}
We present an updated empirical analysis on the ratio of the elastic (integrated) to the total cross section in the c.m. energy interval from 5~GeV to 8~TeV. As in a previous work, we use a suitable analytical parametrization for that ratio (depending on only four free fit parameters) and investigate three asymptotic scenarios: either the black disk limit or scenarios above or below that limit. The dataset includes now the datum at 7~TeV, recently reported by the ATLAS Collaboration. Our analysis favors, once more, a scenario below the black disk, providing  an asymptotic ratio consistent with the rational value $1/3$, namely a gray disk limit. Upper bounds for the ratio of the diffractive (dissociative) to the inelastic cross section are also presented.
\end{abstract}

\vspace{0.3cm}
\centerline{\it XIII International Workshop on Hadron Physics}

\centerline{\it to be published in IOP: Conference Series}
\vspace{0.3cm}

\section{Introduction}

The total cross section ($\sigmatot$) and the elastic cross section ($\sigmael$) are fundamental quantities in the study 
of elastic hadron scattering. Their ratio, as a function of the c.m. energy squared,
\begin{equation}
X(s) \equiv \frac{\sigmael}{\sigmatot}(s),
\end{equation}
plays also an important role \cite{bc,dremin}: it is related to the ratio of the inelastic cross section ($\sigmain$) to the total cross section via unitarity; it is  connected to the central opacity in hadron-hadron collisions (impact parameter representation); it can be related to the ratio between $\sigmatot$ and the elastic slope $B$, which is an important quantity in studies on extensive air showers (cosmic-ray experiments).

However, $X(s)$ connects two quantities associated with soft strong scattering states, which, presently, are not 
directly and unequivocally described from the first principles of QCD. In this situation, a possible strategy of investigation is to obtain phenomenological and/or empirical descriptions of these quantities which may contribute with the
search of suitable calculational schemes in the nonpertubative sector of QCD.

In this communication, we present an empirical approach for studying $X(s)$ in the case of proton-proton ($pp$) and antiproton-proton ($\bar{p}p$) scattering. This analysis follows previous studies by Fagundes and Menon \cite{fm12,fm13} and Fagundes, Menon and Silva \cite{fms15a}. Here, besides the new experimental data on $\sigmatot$ and $\sigmael$ obtained at the CERN-LHC at 7 and 8 TeV by TOTEM Collaboration \cite{totem1,totem2}, we also include the recent ATLAS Collaboration results at 7 TeV \cite{atlas}. 
By means of a suitable empirical parametrization and fits to the experimental data, we investigate different asymptotic scenarios
for $X(s)$. Our analysis favors a scenario below the black-disk limit, consistent with the rational value 1/3.
We discuss some preliminary tests on the selection of free parameters and infer that 1/4 may
constitute an upper bound for the
ratio between the soft diffractive cross section (single/double dissociation) and the inelastic cross section.

\section{Empirical parametrization and asymptotic scenarios}

Our dataset comprises measurements of $\sigmatot$ and $\sigmael$ from $pp$ and $\bar{p}p$ scattering obtained
in accelerator experiments,  from $\sqrt{s_\mathrm{min}} = 5$ GeV up to $\sqrt{s_\mathrm{max}} = 8$ TeV
\cite{pdg14}.
The statistical and systematic uncertainties have been added in quadrature.  The experimental points
for $X$ are shown in figure 1. In this figure we also display the point at 57 TeV obtained via unitarity
from the estimates of  $\sigmatot$ and $\sigmain$ by the Pierre Auger Collaboration \cite{auger}; 
this point did not take part in the data reductions: it is shown only as an illustration of cosmic-ray information
presently available.

\subsection{Empirical parametrization}

On the one hand, from the empirical behavior of $X$ (figure~\ref{f1}), we see that this quantity decreases until $\sqrt{s} \sim 100$ GeV and 
then begins to rise as the energy increases and with a positive curvature. 
On the other hand it is expected that the ratio $X(s)$ becomes constant when $s\to\infty$, i.e. $X(s)$ saturates. 
The value of this constant can change from model to model (see discussion below). 
Since the saturation implies in a negative curvature, it is expected a change of curvature in $X(s)$ 
at some finite energy. Based on these facts and looking for an economical number of free parameters,
we consider the following analytical parametrization:
\begin{equation}
X(s) = A \tanh\{g(s)\},
\end{equation}
with
\begin{equation}
g(s) = \alpha + \beta\ln^\delta(s/s_0) + \gamma\ln(s/s_0),
\end{equation}
where $A$, $\alpha$, $\beta$, $\gamma$ and $\delta$ are dimensionless parameters, in principle, free fit parameters
and $s_0$ is a fixed energy scale.
As constructed, the parameter $A$ corresponds to the asymptotic value of $X(s)$,
$\lim_{s \rightarrow \infty}X(s) = A$
and the hyperbolic tangent takes account of the change of curvature. Therefore, in data
reductions, by fixing $A$ we can define an asymptotic scenario or by letting it free
we can select a scenario.

\subsection{Asymptotic scenarios}
\label{subsec:asymp_scenarios}

The standard scenario in the phenomenological context, typical of eikonal models (for example, \cite{bsw}), concerns the black
disk limit, which predicts $A = 0.5$ \cite{bh}.
However, theoretical, phenomenological and empirical arguments, allow also to consider scenarios
either above or below the black disk, as discussed in what follows.

\subsubsection{Scenarios above the black disk}

From the $s$-channel unitarity,  
\begin{eqnarray}
\frac{\sigmael}{\sigmatot} + \frac{\sigmain}{\sigmatot} = 1.  
\nonumber
\end{eqnarray}
Therefore, a maximum value for the ratio $X$ allowed in the formal context reads
$A = 1$.

Moreover, two well known formal limits (the Froissart-Lukaszuk-Martin bound for total cross section \cite{froissart_martin} and a similar bound for the inelastic cross section \cite{martin}) predict 
\begin{equation}
\sigmatot \leq \frac{\pi}{m_\pi^2} \ln^2 s \quad \mathrm{and} \quad \sigmain \leq \frac{\pi}{4m_\pi^2}\ln^2 s \quad (s\to\infty).
\nonumber
\end{equation}
Now, assuming that \textit{both limits saturate} when $s\to\infty$, it is possible that 
$\sigmain/\sigmatot~\to~1/4$, 
and therefore, from unitarity, we have also a ``formal" possibility
$A = 3/4 = 0.75$ (``formal" result).
We recall that a scenario above the black disk is also indicated in the phenomenological context through the 
$U$-Matrix unitarity scheme \cite{umatrix1}.

\subsubsection{Scenarios below the black disk}

Taking into account the TOTEM empirical fit to $\sigmael$ data \cite{totem1,totem2} 
and the highest-rank fit result to $\sigmatot$ data by the COMPETE Collaboration \cite{compete2},
the central values of the free parameters lead to the asymptotic ratio (see \cite{fms13,ms13a} for details)
$A = 0.436$ (TOTEM/COMPETE).

Moreover, in a recent study on the rise of $\sigmatot$, we have considered a parametrization 
with Regge, Pomeron terms and a leading contribution in the form $\ln^{\gamma} s$, with $\gamma$ either 
fixed to 2 or as a free fit parameter \cite{fms13,ms13a,ms13b}. 
The parametrization was also extended to fit the $\sigmael$ data and in
all cases investigated we have obtained asymptotic ratios $X$ below the
black disk  and consistent with the rational limits between  $1/3$ and $2/5$ \cite{ms13b}. As an extreme
scenario we consider here the lowest value obtained in the analyses (denoted FMS), which,
within the uncertainties from error propagation, reads
$A = 0.30$ (minimum FMS).

In what follows we shall consider the asymptotic limits A = 1.0, 0.75, 0.50, 0.436, 0.30, as representative
values for investigating the three different scenarios. As already commented,
with parametrization (2-3), once $A$ is fixed to each one of the aforementioned values 
we define an asymptotic scenario  and by letting $A$ as a free parameter we can select a scenario.

\section{Fits and results}

Different fits and procedures have been tested in order to select a best value for the energy
scale $s_0$, which led to the energy cutoff as an optimum value. Therefore, we fix $s_0 =$ 25 GeV$^2$.

In what follows, we first present the fits to only $pp$ data, since, as we shall demonstrate, they allow us to fix (eliminate) one 
of the free parameters in (2-3). With this particular parametrization the fits are then extended to 
include $\bar{p}p$ data.

\subsection{Preliminary fits to $pp$ data}

The dataset consists of 29 points (cutoff at 5 GeV), including those at 7 and 8 TeV by the 
TOTEM \cite{totem1,totem2} and ATLAS Collaborations \cite{atlas}. 
We have fixed $A$ to each value referred to above and letting free the parameters $\alpha$, 
$\beta$, $\gamma$ and $\delta$. The results are shown in table~\ref{t1},
together with the statistical information on the fit results: reduced chi-squared 
($\chi^2_\mathrm{red}=\chi^2/\nu$) and integrated probability $P_{\chi^2}=P(\chi^2,\nu)$, where $\nu$ 
is the number of degrees of freedom.

\begin{table}
\caption{\label{t1}Fit results to only $pp$ data through parametrization (2-3) ($\delta$ as a free parameter, 
$A$ fixed) and statistical information ($\nu$ = 25).}
\begin{center}
\begin{tabular}{ccccccc}
\hline
$A$    &     $\alpha$      &      $\beta$       &       $\gamma$     &   $\delta$      & $\chi^2_\mathrm{red}$ & $P_{\chi^2}$\\
fixed  &                   &                    &                    &                       &                       &                         \\
\hline
0.3    & 1.200 $\pm$ 0.046 & -0.604 $\pm$0.048  &  0.240 $\pm$ 0.018 & 0.6494 $\pm$ 0.0039 &          0.846        &        0.684          \\
0.436  &  0.73 $\pm$ 0.16  &  -0.30 $\pm$ 0.11  & 0.0787 $\pm$ 0.050 & 0.52 $\pm$ 0.31   &          0.822        &        0.717            \\
0.5    &  0.62 $\pm$ 0.13  & -0.245 $\pm$0.094  & 0.0625 $\pm$ 0.038 & 0.51 $\pm$ 0.31   &          0.820        &        0.720            \\
0.75   & 0.392 $\pm$ 0.085 & -0.149 $\pm$ 0.63  & 0.0362 $\pm$ 0.022 & 0.50 $\pm$ 0.31   &          0.817        &        0.724            \\
1.0    & 0.289 $\pm$ 0.063 & -0.108 $\pm$ 0.048 &  0.020 $\pm$ 0.016 & 0.49 $\pm$ 0.32   &          0.816        &        0.726            \\
\hline
\end{tabular}
\end{center}
\end{table}

From table~\ref{t1} we note that, with the exception of the extreme case $A = 0.3$, all the results 
for the parameter $\delta$ are compatible with the value 0.5 within the uncertainties.
That suggested us the possibility to fix  $\delta$, reducing, therefore, the number of free parameters.
With that assumption the parametrization (2-3) now reads
\begin{equation}
 X(s) = A \tanh[\alpha + \beta \sqrt{\ln (s/s_0)} + \gamma\ln(s/s_0)],
 \label{eq:par2}
\end{equation}
with $\alpha$, $\beta$, $\gamma$ free parameters ($s_0 = 25$ GeV$^2$). 
We stress the small number of free parameters: 3 in the case of $A$ fixed and
4 if $A$ is let free. 
This parametrization was introduced in \cite{fms15a} and will be employed in what follows.

\subsection{Simultaneous fits to $pp$ and $\bar{p}p$ data}

Now our dataset consists of 42 points. As before, we first consider fits with the fixed
values for $A$, using as initial values for the free parameters those obtained in the
fits to $pp$ data (table~\ref{t1}).

\begin{figure}
\begin{center}
\includegraphics[width=18cm,height=16cm]{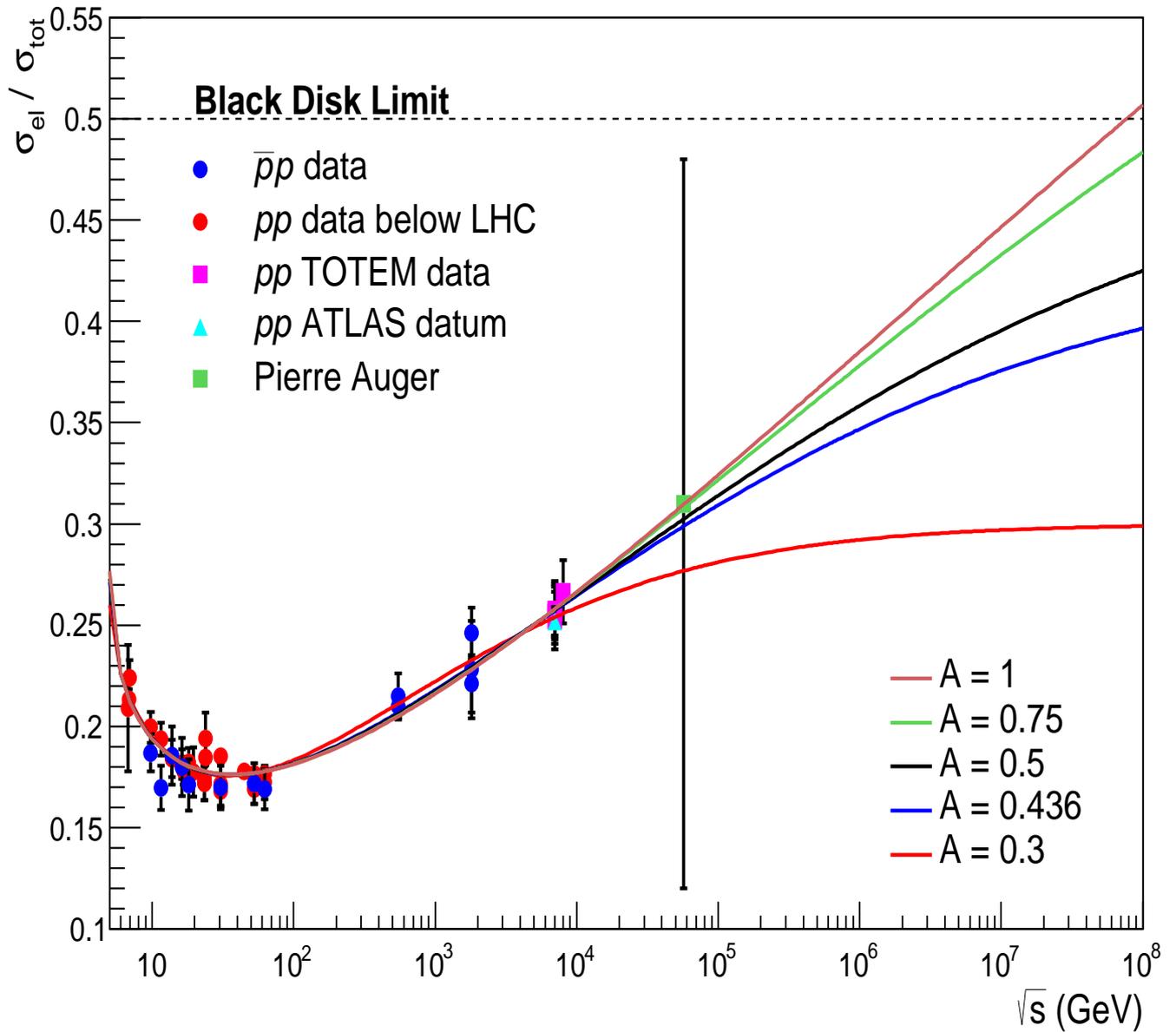}
\end{center}
\caption{\label{f1} Fit results to $pp$ and $\bar{p}p$ data 
on $X = \sigmael/\sigmatot$ through parametrization (5), with
$A$ fixed (parameters from table 2).}
\end{figure}

The fit results and  statistical information are shown in table~\ref{t2} and the corresponding curves together with
the experimental data in figure~\ref{f1}. We see that, despite the distinct asymptotic behaviors, 
all the fit results 
show consistence with the experimental data presently available. In other words, with parametrization (5) 
and $A$ fixed, we can not select an asymptotic scenario.

\begin{table}
\caption{\label{t2} Fit results to $pp$ and $\bar{p}p$ data through parametrization (5) with $A$ fixed.} 
\begin{center}
\begin{tabular}{cccccc}
\hline
$A$   &       $\alpha$      &        $\beta$       &       $\gamma$      & $\chi^2_\mathrm{red}$ & $P_{\chi^2}$\\
fixed &                     &                      &                     &              &               \\
\hline
0.3   & 1.317 $\pm$ 0.053   & -0.658 $\pm$ 0.049   & 0.168 $\pm$ 0.011   &     0.771    & 0.846  \\
0.436 & 0.728 $\pm$ 0.024   & -0.301 $\pm$ 0.021   & 0.0756 $\pm$ 0.0046 &     0.776    & 0.841 \\
0.5   & 0.612 $\pm$ 0.019   & -0.245 $\pm$ 0.017   & 0.0614 $\pm$ 0.0037 &     0.782    & 0.832 \\
0.75  & 0.386 $\pm$ 0.011   & -0.146 $\pm$ 0.010   & 0.0365 $\pm$ 0.0021 &     0.795    & 0.816 \\
1.0   & 0.2846 $\pm$ 0.0083 & -0.1058 $\pm$ 0.0073 & 0.0264 $\pm$ 0.0015 &     0.794    & 0.813 \\
\hline
\end{tabular}
\end{center}
\end{table}

However, a striking result is obtained in the case of $A$ as a free fit parameter.
In fact, despite the strong nonlinearity of the fit, using as initial values of the parameters those 
displayed in table~\ref{t2}, with $A$ free, all the data reductions converge to the same result, given by

\begin{eqnarray}
A =  0.332 \pm 0.049, \quad \alpha = 1.09 \pm 0.28, \quad  \beta = -0.51 \pm 0.18, \quad \gamma = 0.129 \pm 0.046 \\
\chi^2/\nu = 0.783, \quad P(\chi^2,\nu) = 0.829 \qquad \qquad \qquad \qquad \qquad \nonumber
\end{eqnarray}

\begin{figure}
\begin{center}
\includegraphics[width=18cm,height=16cm]{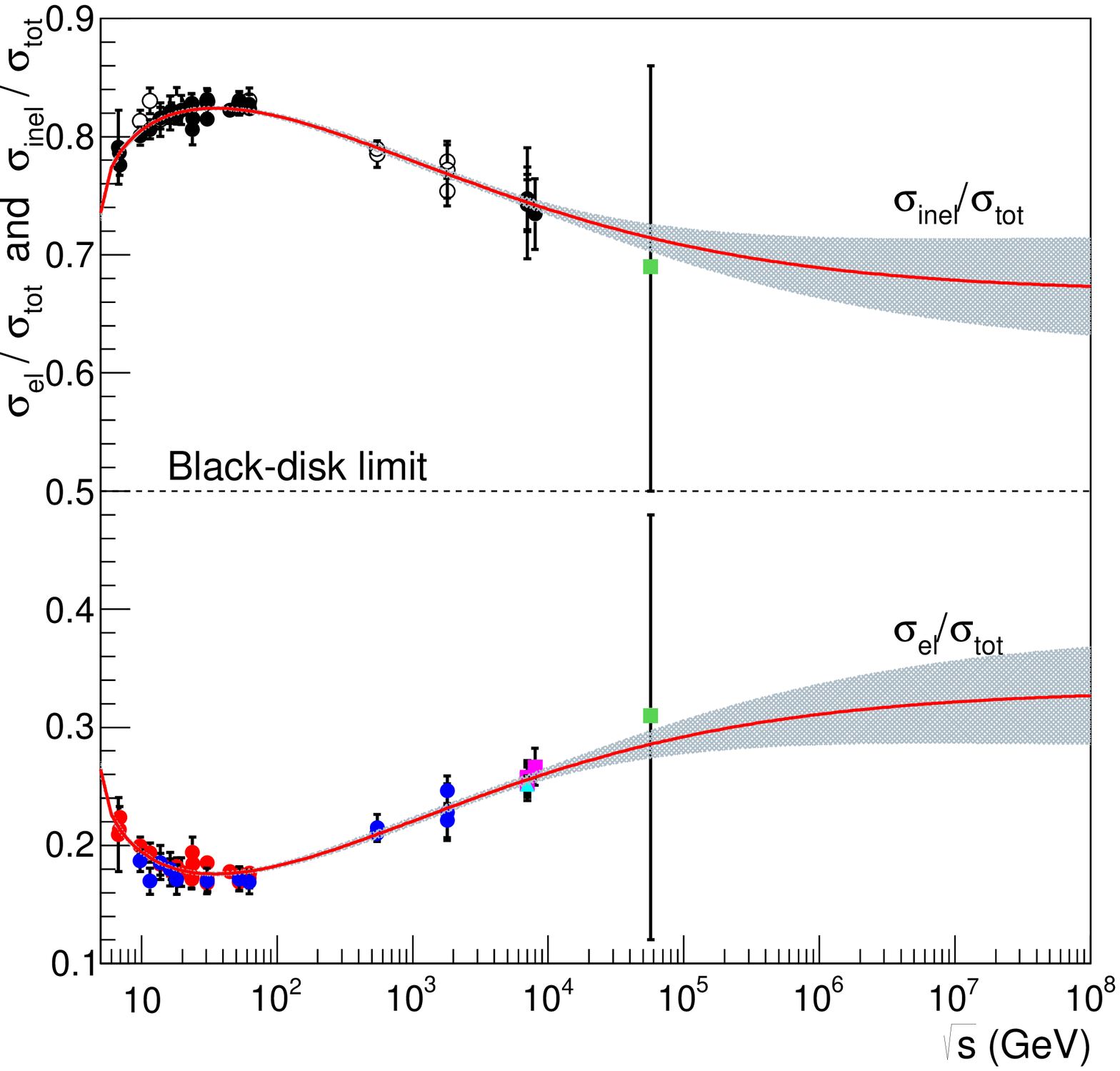}
\end{center}
\caption{\label{f2} 
Fit result to $X = \sigmael/\sigmatot$ through parametrization (5) and $A$ free parameter (parameters (6)) and \textit{prediction} 
for the ratio $\sigmain/\sigmatot$ (via unitarity), together with their corresponding uncertainty regions.}
\end{figure}

The comparison with the experimental data is shown in figure~\ref{f2}, where the  uncertainty region 
has been evaluated through error propagation from the fit parameters (variances and covariances).
Also shown in this figure is the prediction for the ratio $\sigmain/\sigmatot$, obtained via unitarity,
together with the experimental data. Numerical predictions for these ratios at some energies
of interest are displayed in table~\ref{t3} (second and third columns).

We conclude that our analysis favors a scenario below that of the black disk, with
asymptotic value compatible with the rational limit
\begin{eqnarray}
A = 0.332 \pm 0.049 \to 1/3.
\nonumber
\end{eqnarray}

We note that this result is not in conflict with the Pumplin bound \cite{pumplin},
\begin{eqnarray}
\frac{\sigmael}{\sigmatot} + \frac{\sigmadiff}{\sigmatot} \leq 1/2,
\nonumber
\end{eqnarray}
where $\sigmadiff$ stands for the soft diffractive cross section (single and double dissociation).
If we consider the saturation of the Pumplin bound (equality in the above equation) yet at the
LHC energy region \cite{lipari},  we can infer the predictions for the ratio $\sigmadiff/\sigmatot$ 
displayed in table~\ref{t3} (last column). 
Moreover, using the Pumplin bound and Unitarity, we can also infer an upper bound for ratio $\sigmadiff/\sigmain$,
namely
\begin{eqnarray}
R(s) \equiv \frac{\sigmadiff}{\sigmain} \leq \frac{1/2 - X(s)}{1-X(s)}.
\nonumber
\end{eqnarray}
In this case, the maximum asymptotic value for $R$ is also compatible
with a rational number:
\begin{eqnarray}
R_\mathrm{max} = 0.251 \pm 0.055\ \to \ 1/4 \quad \mathrm{as} \quad s\to\infty.
\nonumber
\end{eqnarray}

\begin{table}
\caption{\label{t3} Predictions for different ratios at the LHC energy region.} 
\begin{center}
\begin{tabular}{cccc}
\hline
$\sqrt{s}$ (TeV) & $\sigmael/\sigmatot$ & $\sigmain/\sigmatot$ & $\sigmadiff/\sigmatot$ \\
\hline
   2.76      &    0.2394 $\pm$ 0.0036     &    0.7606 $\pm$ 0.0036  &  0.2606 $\pm$ 0.0036 \\
   7         &    0.2556 $\pm$ 0.0043     &    0.7444 $\pm$ 0.0043  &  0.2444 $\pm$ 0.0043 \\
   8         &    0.2578 $\pm$ 0.0046     &    0.7422 $\pm$ 0.0046  &  0.2422 $\pm$ 0.0046 \\
   13        &    0.2655 $\pm$ 0.0058     &    0.7345 $\pm$ 0.0058  &  0.2345 $\pm$ 0.0058 \\
   14        &    0.2666 $\pm$ 0.0061     &    0.7334 $\pm$ 0.0061  &  0.2334 $\pm$ 0.0061 \\
\hline
\end{tabular}
\end{center}
\end{table}

\section{Conclusions and final remarks}

We have presented an empirical analysis of the experimental data presently
available on the ratio $X=\sigmael/\sigmatot$ above 5 GeV, including all the recent
data at 7 and 8 TeV obtained at the LHC by the TOTEM and ATLAS Collaborations.
Our model-independent parametrization, characterized by a small number of free parameters
(3 or 4), allows consistent descriptions of all the experimental data analyzed
(figures~\ref{f1} and \ref{f2}).

By letting free the parameter $A$, the resulting asymptotic value of $X$  is compatible with a scenario \textit{below} 
the black-disk limit, namely $0.332 \pm 0.049 \rightarrow 1/3$. This value is in agreement with our previous 
results based in distinct analyses
\cite{fms13,ms13a,ms13b}; it is also in plenty agreement with the recent phenomenological
approach by Kohara, Ferreira and Kodama \cite{kohara_ferreira_kodama} which predicts $X\to 1/3$ for $s\to\infty$.
Taking into account the Pumplin bound, it was possible to infer an upper bound for the ratio $\sigmadiff/\sigmain$,
consistent with the rational value 1/4.

Our next steps concern a review on all the results presented here and  in \cite{fm12,fm13,fms15a}, 
extensions of the analysis treating other choices for $g(s)$
and detailed investigation on the change of curvature in $X(s)$ at the LHC energy region \cite{fms15c}.

\section*{Acknowledgments}

Research supported by FAPESP-CAPES, Contract 2014/00337-8 (DAF) and
FAPESP, Contract 2013/27060-3 (PVRGS).


\end{document}